\documentclass[a4paper]{article}
  \usepackage[margin=3cm]{geometry}
  \usepackage{amsthm}
  
  \newtheorem{theorem}{Theorem}
  \newtheorem*{theorem*}{Theorem}
  \newtheorem{question}[theorem]{Question}
\usepackage[utf8]{inputenc}
\usepackage[T1]{fontenc}
\usepackage{graphicx}
\usepackage{grffile}
\usepackage{longtable}
\usepackage{wrapfig}
\usepackage{rotating}
\usepackage[normalem]{ulem}
\usepackage{amsmath}
\usepackage{textcomp}
\usepackage{amssymb}
\usepackage{capt-of}
\usepackage{hyperref}
\newcommand{\U}{\mathcal{U}}
\newcommand{\tuple}[1]{\langle #1 \rangle}
\newcommand{\syntconst}[1]{\mathsf{#1}}
\newcommand{\syntconstn}[1]{\mathbf{#1}}
\newcommand{\syntconstb}[1]{\mathbb{#1}}
\newcommand{\theory}[1]{\mathbb{#1}}
\newcommand{\type}{\syntconst{type}}
\newcommand{\rel}{\syntconst{rel}}
\newcommand{\ind}{\syntconst{ind}}
\newcommand{\rec}{\syntconst{rec}}
\newcommand{\proj}{\syntconst{pr}}
\newcommand{\refl}{\syntconst{refl}}
\newcommand{\id}{\syntconst{id}}
\newcommand{\concat}{\cdot}
\newcommand{\whisk}[1]{\concat_{\syntconst{#1}}}
\newcommand{\transport}{\syntconst{transport}}
\newcommand{\apply}{\syntconst{ap}}

\newcommand{\base}{\syntconst{base}}
\newcommand{\Sphloop}{\syntconst{loop}}
\newcommand{\W}{\syntconst{W}}
\newcommand{\Wsup}{\syntconst{sup}}
\newcommand{\inl}{\syntconst{inl}}
\newcommand{\inr}{\syntconst{inr}}
\newcommand{\glue}{\syntconst{glue}}
\newcommand{\suspension}{\Sigma}
\newcommand{\suspN}{\syntconst{N}}
\newcommand{\suspS}{\syntconst{S}}
\newcommand{\join}{\star}
\newcommand{\Zero}{\syntconstn{0}}
\newcommand{\One}{\syntconstn{1}}
\newcommand{\Two}{\syntconstn{2}}
\newcommand{\N}{\syntconstb{N}}
\newcommand{\Sph}{\syntconstb{S}}
\newcommand{\point}{\bullet}
\newcommand{\Loop}{\Omega}
\newcommand{\relation}[1]{\mathcal{#1}}
\newcommand{\related}[1]{\bar{#1}}
\newcommand{\Rel}{\mathcal{R}el}
\newcommand{\itpr}[1]{[\![#1]\!]}
\newcommand{\LEM}{\syntconst{LEM}}
\newcommand{\isequiv}{\syntconst{isequiv}}
\newcommand{\idtoequiv}{\syntconst{idtoequiv}}
\newcommand{\ua}{\syntconst{ua}}
\usepackage{tikz}
\usetikzlibrary{cd}
\author{Taichi Uemura}
\date{\today}
\title{Homotopies for Free!}
\hypersetup{
 pdfauthor={Taichi Uemura},
 pdftitle={Homotopies for Free!},
 pdfkeywords={},
 pdfsubject={},
 pdfcreator={Emacs 25.1.1 (Org mode 9.0.5)}, 
 pdflang={English}}
\begin{document}

\maketitle
\begin{abstract}
We show ``free theorems'' in the style of Wadler
for polymorphic functions in homotopy type theory
as consequences of the abstraction theorem.
As an application, it follows that
every space defined as a higher inductive type
has the same homotopy groups as some type of polymorphic functions
defined without univalence or higher inductive types.
\end{abstract}

\section{Introduction}
\label{sec:orgab9361f}
Given a closed term of type of polymorphic functions
defined in homotopy type theory \cite{hottbook},
we can derive a theorem that it satisfies.
For example,
let \(t\) be a closed term of type
\[t : \prod_{X : \U}\prod_{x : X}x = x \to x = x.\]
Then we have a theorem
\[\prod_{X, X' : \U}\prod_{f : X \to X'}
  \prod_{x : X}\prod_{p : x = x}t(fp) = f(tp)\]
in homotopy type theory, in the sense that
there is a closed term of this type.

Such theorems are ``free theorems'' in the style of Wadler \cite{wadler1989free}
for homotopy type theory.
Original free theorems for polymorphic type theory
are consequences of relational parametricity \cite{reynolds1983types}
and have a lot of applications including
short cut fusion \cite{gill1993short,svenningsson2002shortcut,takano1995shortcut},
non-definability of polymorphic equality \cite{wadler1989free},
and encoding initial algebras and final coalgebras
in pure polymorphic lambda calculus \cite{hasegawa1994categorical}.
Recently relational parametricity and free theorems for dependent type theory
have been studied by several authors.
Atkey et al. \cite{atkey2014relationally}
constructed relationally parametric models of Martin-L\"{o}f type theory
and proved a simple free theorem
and the existence of initial algebras for indexed functors.
Takeuti \cite{takeuti2001parametricity} studied relational parametricity for the lambda cube
and proved adjoint functor theorem internally.
Bernardy et al. \cite{bernardy2012proofs}
studied relational parametricity for pure type systems
and free theorems for dependently typed functions.

In this paper we show free theorems specific to homotopy type theory
such as the example given in the first paragraph
where the type \(\prod_{X : \U}\prod_{x : X}x = x \to x = x\)
seems to be trivial without homotopy-theoretic interpretation.
A difference between free theorems for homotopy type theory
and original free theorems for polymorphic type theory is that
in homotopy type theory they are represented by \emph{homotopies}
instead of \emph{equalities}.
This difference causes some problems
related to proof-relevance and higher dimensional homotopies.
One approach to these problems is \emph{higher dimensional parametricity}
\cite{ghani2016proof,johann2017cubical}
and to state free theorems as \emph{coherent homotopies}.
Both in \cite{ghani2016proof} and \cite{johann2017cubical},
the target languages are polymorphic lambda calculus
which does not have higher dimensional structures.
On the other hand, our target language, homotopy type theory,
has already higher dimensional structures,
and thus ordinary free theorems for higher dimensional types work well.

To explain this, let us see an example.
Consider a canonical embedding
\begin{align*}
i &: A \to \prod_{X : \U}(A \to X) \to X &
i &\equiv \lambda a. \lambda (X, g).ga
\end{align*}
for a base type \(A : \U\).
In polymorphic type theory
it follows from a free theorem that \(i\) is an isomorphism.
In homotopy type theory
an immediate consequence of a free theorem is the fact that
\(i\) is \emph{\(0\)-connected}, that is, it induces
a bijection between the sets of connected components.
A \(0\)-connected map is far from an isomorphism.
However, for each \(n \ge 1\) and \(a : A\),
it follows from a free theorem for the type
\[\prod_{X : \U}\prod_{g : A \to X}\Loop^{n}(X, ga)\]
that \(i\) induces a \(0\)-connected map
\[\Loop^{n}(i) : \Loop^{n}(A, a) \to
  \Loop^{n}(\prod_{X : \U}(A \to X) \to X, ia)\]
between the \(n\)-th loop spaces.
Therefore we conclude that \(i\) is \emph{\(\infty\)-connected}, that is,
it induces a bijection between the \(n\)-th homotopy groups
for each \(n \ge 0\).
Hence the types \(A\) and \(\prod_{X : \U}(A \to X) \to X\)
are equivalent from homotopical point of view.

For a concrete (higher) inductive type \(A\),
the type \(\prod_{X : \U}(A \to X) \to X\)
is equivalent to a type definable in Martin-L\"{o}f type theory \cite{martin-lof1975intuitionistic}
without univalence or higher inductive types.
For example,
\[(\prod_{X : \U}(\Sph^{n} \to X) \to X) \simeq
  (\prod_{X : \U}\prod_{x : X}\Loop^{n}(X, x) \to X)\]
where \(\Sph^{n}\) is the \(n\)-dimensional sphere.
The right hand side of this equivalence
is the \emph{Church encoding} of \(n\)-sphere,
proposed by Shulman\footnote{\url{https://homotopytypetheory.org/2011/04/25/higher-inductive-types-via-impredicative-polymorphism/}}.
It follows from the previous paragraph that
every space can be identified via an \(\infty\)-connected map
with its Church encoding.
The Church encoding of a space suggests that
generators of its homotopy groups are definable without univalence or higher inductive types.
For example the generator of \(\pi_{3}(\Sph^{2})\)
can be defined as polymorphic functions of type
\(\prod_{X : \U}\prod_{x : X}\Loop^{2}(X, x) \to \Loop^{3}(X, x)\).
We can say that the univalence axiom and higher inductive types
are used only for \emph{proving} that \(\pi_{3}(\Sph^{2})\) is the integers
but not needed for \emph{creating} the generator of \(\pi_{3}(\Sph^{2})\).

Free theorems for general open terms in homotopy type theory
should follow from relational parametricity,
but it seems to be hard to axiomatize relational parametricity for homotopy type theory.
Thus we focus on free theorems for closed terms
as the first step to understanding relational parametricity
for homotopy type theory,
because free theorems for closed terms follow from Reynolds's \emph{abstraction theorem} \cite{reynolds1983types}
without any assumptions.
Informally, it says that terms evaluated under related environments yield related values.
We show the abstraction theorem for homotopy type theory
via a syntactic transformation of a term in homotopy type theory to another.
The key to prove the abstraction theorem is the fact that
binary type families in homotopy type theory form a model of homotopy type theory
which we call the \emph{relational model}.
Then the abstraction theorem is the soundness of the interpretation
of types as binary type families.
There is a category-theoretic proof of this fact
using Shulman's inverse diagrams of type-theoretic fibration categories \cite{shulman2015inverse}
or fibred type-theoretic fibration categories introduced by the author \cite{uemura2017fibred}.
In this paper we give a syntactic proof
in order to make the paper self-contained.
We also show a new result on inductive data types:
for a type theory with \emph{indexed \(\W\)-types},
originally called \emph{general trees} \cite{nordstrom1990programming,petersson1989set},
the relational model has indexed \(\W\)-types.
The construction of indexed \(\W\)-types in the relational model
is essentially same as that of \(\W\)-types in the gluing construction
for a cartesian functor between \(\Pi\W\)-pretoposes
\cite{moerdijk2000wellfounded,vandenberg2006predicative}.

The study of relational parametricity via syntactic transformations is not new.
Abadi et al. \cite{abadi1993formal} and
Plotkin and Abadi \cite{plotkin1993logic}
introduced logic for parametricity
where the abstraction theorem is the soundness of
the interpretations of terms in System F as proofs in their logic.
Wadler pointed out that
Reynolds's abstraction theorem can be seen as
a transformation of a term in System F to a proof in second-order logic
\cite{wadler2003girard,wadler2007girard}.
Takeuti \cite{takeuti2001parametricity} and
Bernardy et al. \cite{bernardy2012proofs}
studied relational parametricity for the lambda cube
and pure type systems respectively
via syntactic transformations of a term in one type theory to another.
Since homotopy type theory, even Martin-L\"{o}f type theory,
is powerful enough to express predicates
(\emph{reflective} in terms of \cite{bernardy2012proofs}),
we can transform a term in homotopy type theory
to another in homotopy type theory itself.
Our contribution is to give transformations of
identity types, the univalence axiom and some higher inductive types.

\textbf{Organization.}
We begin in Section \ref{orgab5e494}
by recalling some important types and functions in homotopy type theory.
Section \ref{org72cd84a} and \ref{orgea7ae87}
are the core of this paper.
In Section \ref{org72cd84a}
we explain what the abstraction theorem is.
In Section \ref{orgea7ae87},
we give some free theorems as corollaries of the abstraction theorem.
In Section \ref{orgbf2ea29},
we discuss Church encodings of higher inductive types
and give the generator of \(\pi_{3}(\Sph^{2})\)
as a polymorphic function.
We prove the abstraction theorem
in Section \ref{org2b91024}, \ref{org103b3d7} and \ref{org8cf691c}.

\section{Preliminaries on Homotopy Type Theory}
\label{sec:org766c824}
\label{orgab5e494}
We recall some types and functions in homotopy type theory
which are used in Section \ref{org72cd84a} and \ref{orgea7ae87}.
See \cite{hottbook} for details.

The key idea of homotopy type theory is to identify types as spaces,
elements as points and equalities as \emph{paths}.
We think of an identity type \(x : A, y : A \vdash x = y \ \type\)
as the \emph{space of paths from \(x\) to \(y\)}.
Under this identification,
reflexivity, transitivity and symmetry
correspond to \emph{constant path} \(\refl_{x} : x = x\),
\emph{path concatenation} \((-) \concat (-) : x = y \to y = z \to x = z\)
and \emph{path inversion} \((-)^{-1} : x = y \to y = x\) respectively.
A function \(f : A \to B\) acts on paths as
\(\apply(f, -) : x = y \to fx = fy\) for all \(x, y : A\),
and we will often write \(\apply(f, p)\) as \(fp\) for \(p : x = y\).
Corresponding to indiscernability of identicals,
there is a function \(\transport^{C}(p, -) : C(x) \to C(y)\)
for \(x : A \vdash C(x) \ \type\), \(x, y : A\) and \(p : x = y\).
Since the symbol ``\(=\)'' is reserved for identity types,
we write \(a \equiv b\) when expressions \(a\) and \(b\)
are \emph{judgmentally} or \emph{definitionally} equal.

A function \(f : A \to B\) also acts on higher dimensional paths.
For \(x_{1}, y_{1} : A\), \(x_{2}, y_{2} : x_{1} = y_{1}\),
\dots{}, \(x_{n}, y_{n} : x_{n - 1} = y_{n - 1}\),
we define \(\apply_{n}(f, -) : x_{n} = y_{n} \to \apply_{n - 1}(f, x_{n}) = \apply_{n - 1}(f, y_{n})\)
as \(\apply_{0}(f, z) \equiv fz\) and
\(\apply_{n}(f, p) \equiv \apply(\apply_{n - 1}(f, -), p)\).
We often write \(\apply_{n}(f, p)\) as \(fp\).
There are compositions of higher dimensional paths.
For \(x_{0}, y_{0} : A\), \(x_{1}, y_{1} : x_{0} = y_{0}\),
\dots{}, \(x_{n}, y_{n} : x_{n - 1} = y_{n - 1}\), \(\sigma : x_{n} = y_{n}\),
\(p : x' = x_{0}\) and \(q : y_{0} = y'\),
we set \(p \whisk{l} \sigma \equiv \apply_{n}(\lambda s.p \concat s, \sigma)\)
and \(\sigma \whisk{r} q \equiv \apply_{n}(\lambda s.s \concat q, \sigma)\).
These operations \(\whisk{l}\) and \(\whisk{r}\) are called \emph{whiskering}.

A \emph{pointed type} is a pair \((A, a)\)
of type \(A\) and its inhabitant \(a : A\) called a \emph{base point}.
For a pointed type \((A, a)\) and a natural number \(n \ge 0\),
the \emph{\(n\)-th loop space \(\Loop^{n}(A, a)\) of \(A\) at \(a\)}
is a pointed type defined inductively as
\(\Loop^{0}(A, a) \equiv (A, a)\) and
\(\Loop^{n + 1}(A, a) \equiv \Loop^{n}(a = a, \refl_{a})\).
Write \(\refl^{n}_{a} : \Loop^{n}(A, a)\) for the base point of \(\Loop^{n}(A, a)\).
A function \(f : A \to B\) acts on loop spaces
as \(\apply_{n}(f, -) : \Loop^{n}(A, a) \to \Loop^{n}(B, fa)\).

A path space of a product space \(A \times B\)
is a product of path spaces:
\((\tuple{a, b} = \tuple{a', b'}) \simeq (a = a') \times (b = b')\)
for \(a, a' : A\) and \(b, b' : B\).
We think of a pair \(\tuple{p, q}\) of paths \(p : a = a'\) and \(q : b = b'\)
as a path \(\tuple{a, b} = \tuple{a', b'}\) in \(A \times B\).
Similarly, we regard a pair \(\tuple{l, k}\)
of \(n\)-loops \(l : \Loop^{n}(A, a)\) and \(k : \Loop^{n}(B, b)\)
as an \(n\)-loop in \(A \times B\) at \(\tuple{a, b}\).

Let \(x : A \vdash B(x) \ \type\) be a type family.
For a path \(p : a = a'\) in \(A\) and points \(b : B(a)\) and \(b' : B(a')\),
the \emph{path space from \(b\) to \(b'\) over \(p\)},
written \(b =_{p} b'\), is the type \(\transport^{B}(p, b) = b'\).
For an \(n\)-loop \(l : \Loop^{n}(A, a)\) and a point \(b : B(a)\),
the \emph{\(n\)-th loop space of \(B\) at \(b\) over \(l\)},
written \(\Loop_{l}^{n}(B, b)\),
is the type \(\apply_{n - 1}(\lambda p.\transport^{B}(p, b), l) = \refl^{n - 1}_{b}\).

\section{Abstraction Theorem Explained}
\label{sec:org2f44ed5}
\label{org72cd84a}
The abstraction theorem for polymorphic type theory
is explained in terms of set-theoretic relations.
For dependent type theory,
we use type-theoretic relations, namely binary type families.

For a binary type family \(x : A, x' : A' \vdash \relation{A}(x, x') \ \type\),
a \emph{family on \(\relation{A}\)} is a triple of
\(x : A \vdash B(x) \ \type\), \(x' : A' \vdash B'(x') \ \type\) and
\(x : A, x' : A', \related{x} : \relation{A}(x, x'), y : B(x), y' : B'(x')
  \vdash \relation{B}(\related{x}, y, y') \ \type\),
written \(\related{x} : \relation{A} \vdash \relation{B}(\related{x}) \ \rel\)
in short.
Note that \(\relation{B}\) depends on \(x : A\) and \(x' : A'\) implicitly.

Let \(\related{x} : \relation{A} \vdash \relation{B}(\related{x}) \ \rel\)
be a family on a binary type family \(\relation{A}\).
The \emph{dependent product of \(\relation{B}\) over \(\relation{A}\)} is the binary type family
\[f : \prod_{x : A}B(x), f' : \prod_{x' : A'}B'(x') \vdash
  \prod_{x : A}\prod_{x' : A'}\prod_{\related{x} : \relation{A}(x, x')}\relation{B}(\related{x}, fx, f'x') \ \type.\]
The \emph{dependent sum of \(\relation{B}\) over \(\relation{A}\)} is the binary type family
\[z : \sum_{x : A}B(x), z' : \sum_{x' : A'}B'(x') \vdash
  \sum_{\related{x} : \relation{A}(\proj_{1}(z), \proj_{1}(z'))}\relation{B}(\related{x}, \proj_{2}(z), \proj_{2}(z')) \ \type.\]

For a binary type family \(\relation{A}\),
the \emph{path space of \(\relation{A}\)} is the family
\begin{align*}
& x_{0} : A, x'_{0} : A', \related{x}_{0} : \relation{A}(x_{0}, x'_{0}),
x_{1} : A, x'_{1} : A', \related{x}_{1} : \relation{A}(x_{1}, x'_{1}), \\
& p : x_{0} = x_{1}, p' : x'_{0} = x'_{1} \vdash \related{x}_{0} =_{\tuple{p, p'}} \related{x}_{1} \ \type
\end{align*}
on two copies of \(\relation{A}\).

A \emph{universe} of binary type families is a binary type family
\[X : \U, X' : \U \vdash X \to X' \to \U \ \type\]
where \(\vdash \U \ \type\) is a universe of types.

For each type constant \(C\) (for example, \(\Zero\), \(\One\), \(\Two\),
\(\N\), \(\Sph^{1}\), \(\Sph^{2}\) and so on),
we associate it with a binary type family
\[c : C, c' : C \vdash c = c' \ \type.\]
Then, by induction, we can associate each type family \(x : X \vdash A(x) \ \type\)
with a family of binary type families
\[x : X, x' : X, \related{x} : \itpr{X}(x, x'), a : A(x),
  a' : A(x') \vdash \itpr{A}(\related{x}, a, a') \ \type.\]
Now the abstraction theorem can be described as follows:

\begin{theorem*}[Abstraction Theorem]
For each term \(x : X \vdash t(x) : A(x)\),
there exists a term
\[x : X, x' : X', \related{x} : \itpr{X}(x, x') \vdash
  \hat{t}(\related{x}) : \itpr{A}(\related{x}, t(x), t(x')).\]
In particular, for each closed term \(\vdash t : A\),
there exists a closed term
\[\vdash \hat{t} : \itpr{A}(t, t).\]
\end{theorem*}

\section{Abstraction Theorem Applied}
\label{sec:org9ff6e9d}
\label{orgea7ae87}

\subsection{Concatenation of a Loop}
\label{sec:org6887b39}
\label{orgd10ced1}
Let \(t\) be a closed term of type
\[t : \prod_{X : \U}\prod_{x : X}x = x \to x = x.\]
One might guess that \(t\) is an iterated concatenation of a loop,
that is,
\[t(p) \equiv \underbrace{p \concat \dots \concat p}_{\text{\(n\) times}}\]
for a fixed integer \(n\),
where negative \(n\) means \((-n)\) times concatenation of the inversion of \(p\).
In fact any closed term of this type
must be homotopic to some iterated concatenation of a loop,
but one can derive a theorem without this fact.
We show that the type
\[\prod_{X, X' : \U}\prod_{f : X \to X'}
   \prod_{x : X}\prod_{p : x = x}t(fp) = f(tp)\]
is inhabited.
From the abstraction theorem we have a closed term
\[\hat{t} : \prod_{(X : \U, X' : \U, \relation{X} : X \to X' \to \U)}
   \prod_{(x : X, x' : X', \related{x} : \relation{X}(x, x'))}
   \prod_{(p : x = x, p' : x' = x', \related{p} : \related{x} =_{\tuple{p, p'}} \related{x})}
   \related{x} =_{\tuple{tp, tp'}} \related{x}.\]
For a function \(f : X \to X'\) of \(\U\)-small types,
let \(\relation{X}(x, x') \equiv fx = x'\).
One can prove that, for \(p : x = x\), \(p' : x' = x'\) and \(\related{x} : fx = x'\),
the type \(\related{x} =_{\tuple{p, p'}} \related{x}\) is equivalent to
the type \(\related{x} \concat p' = fp \concat \related{x}\).
Letting \(x' \equiv fx\) and \(\related{x} \equiv \refl_{fx}\),
we have an inhabitant of the type
\[\prod_{p : x = x, p' : fx = fx, \related{p} : p' = fp}
   tp' = f(tp).\]
Finally we set \(p' \equiv fp\) and \(\related{p} \equiv \refl_{fp}\).
Then we have an inhabitant of the type
\[\prod_{x : X}\prod_{p : x = x}t(fp) = f(tp).\]

\subsection{Loop Operations}
\label{sec:orgbe21050}
\label{orgb86f605}
The example in Section \ref{orgd10ced1} can be generalized.
Let \(n\) and \(k\) be natural numbers
and \(t\) a closed term of type
\[t : \prod_{X : \U}\prod_{x : X}\Loop^{n}(X, x) \to \Loop^{k}(X, x).\]
The example in Section \ref{orgd10ced1}
is the case when \(n = k = 1\).
This type represents the \(k\)-th loop space of \(n\)-sphere,
discussed in Section \ref{orgbf2ea29},
and thus we could not guess what function \(t\) is.
However, we can derive a theorem about \(t\).
We show that the type
\[\prod_{X, X' : \U}\prod_{f : X \to X'}
   \prod_{x : X}\prod_{p : \Loop^{n}(X, x)}
   t(fp) = f(tp)\]
is inhabited.
From the abstraction theorem we have a closed term
\[\hat{t} : \prod_{(X : \U, X' : \U, \relation{X} : X \to X' \to \U)}
   \prod_{(x : X, x' : X', \related{x} : \relation{X}(x, x'))}
   \prod_{(p : \Loop^{n}(X, x), p' : \Loop^{n}(X, x'), \related{p} : \Loop^{n}_{\tuple{p, p'}}(\relation{X}, \related{x}))}
   \Loop^{k}_{\tuple{tp, tp'}}(\relation{X}, \related{x}).\]
For a function \(f : X \to X'\) of \(\U\)-small types,
let \(\relation{X}(x, x') \equiv fx = x'\).
One can prove that, for \(p : \Loop^{n}(X, x)\),
\(p' : \Loop^{n}(X, x')\) and \(\related{x} : fx = x'\),
the type \(\Loop^{n}_{\tuple{p, p'}}(\relation{X}, \related{x})\) is equivalent to
the type \(\related{x} \whisk{l} p' = fp \whisk{r} \related{x}\).
Letting \(x' \equiv fx\), \(\related{x} \equiv \refl_{fx}\),
\(p' \equiv fp\) and \(\related{p} \equiv \refl_{fp}\),
we have an inhabitant of the type
\(\prod_{x : X}\prod_{p : \Loop^{n}(X, x)}t(fp) = f(tp)\).

\subsection{Action on Loops}
\label{sec:org23d303d}
Let \(t\) be a closed term of type
\[t : \prod_{X, Y : \U}\prod_{f : X \to Y}\prod_{x : X}x = x \to fx = fx.\]
One might guess that \(t(f, p) \equiv \apply(f, p)\).
Of course, \(t\) could be another function, for example,
\(t(f, p) \equiv \apply(f, p \concat p)\).
However, intuitively only \(\apply(f, p)\)
is an interesting function of this type,
because \(\apply(f, p \concat p)\) is a composition of \(\apply(f, p)\)
and a loop concatenation, and the latter does not use \(f\).

Let us to formulate this intuition.
We show that the type
\[\prod_{X, Y : \U}\prod_{f : X \to Y}\prod_{x : X}\prod_{p : x = x}
   t(f, p) = f(t(\id_{X}, p))\]
is inhabited.
This means that, for any \(t\),
\(t(f, -)\) is a composition of \(\apply(f, -)\)
after a loop operation \(t(\id_{X}, -) : x = x \to x = x\).
From the abstraction theorem we have an inhabitant of the type
\begin{align*}
& \prod_{X', X, Y', Y : \U}\prod_{g : X' \to X}\prod_{h : Y' \to Y}
\prod_{f' : X' \to Y'}\prod_{f : X \to Y}
\prod_{\sigma : \prod_{x' : X'}f(gx') = h(f'x')} \\
& \prod_{x' : X'}\prod_{p' : x' = x'}
t(f, gp') \concat \sigma(x') = \sigma(x') \concat h(t(f', p')).
\end{align*}
Letting \(X' \equiv Y' \equiv X\), \(h \equiv f\),
\(f' \equiv g \equiv \id_{X}\) and \(\sigma \equiv \lambda x.\refl_{fx}\),
we have
\[\prod_{X, Y : \U}\prod_{f : X \to Y}\prod_{x : X}\prod_{p : x = x}
   t(f, p) = f(t(\id_{X}, p)).\]
Note that, from Section \ref{orgd10ced1},
we also have \(f(t(\id_{X}, p)) = t(\id_{Y}, fp)\).

\subsection{An Embedding}
\label{sec:orgfba250a}
\label{org6d38184}
For a base type \(A : \U\) such as \(\N\) and \(\Sph^{1}\),
let \(\widetilde{A} \equiv \prod_{X : \U}(A \to X) \to X\).
There are back and forth functions between the types
\(A\) and \(\widetilde{A}\) as follows:
\begin{align*}
i & : A \to \prod_{X : \U}(A \to X) \to X &
i & \equiv \lambda(a : A).\lambda(X : \U, g : A \to X).ga \\
j & : (\prod_{X : \U}(A \to X) \to X) \to A &
j & \equiv \lambda(\varphi : \prod_{X : \U}(A \to X) \to X).\varphi_{A}(\id_{A}).
\end{align*}
Clearly \(j \circ i \equiv \id\),
but \(i \circ j \equiv \id\) or even \(i \circ j \sim \id\)
does not hold.
However, given a closed term \(t : \widetilde{A}\),
we can construct a closed term of type
\[\prod_{X : \U}\prod_{g : A \to X}(i(jt))g = tg.\]
To show this, let \(t : \widetilde{A}\) be a closed term.
From the abstraction theorem we can get a closed term of type
\(\prod_{X_{0}, X : \U}\prod_{f : X_{0} \to X}
   \prod_{g : A \to X_{0}}f(tg) = t(f \circ g)\).
Taking \(X_{0} \equiv A\) and \(g \equiv \id_{A}\),
we get an inhabitant of the type \(f(t(\id_{A})) = t(f)\).
Now, for \(X : \U\) and \(g : A \to X\), we have
\((i(jt))g = g(jt) = g(t(\id_{A})) = t(g).\)

\subsection{An \(\infty\)-Connected Map}
\label{sec:org55b1471}
\label{org8f38bec}
For a type \(A\), let \(\pi_{0}(A)\) be the set of
homotopy equivalence classes of closed terms of \(A\)
which we call the \emph{\(0\)-th homotopy group of \(A\)}.
For a point \(a : A\) and a natural number \(n\),
the \emph{\(n\)-the homotopy group of \(A\) at \(a\)}, written \(\pi_{n}(A, a)\),
is the set \(\pi_{0}(\Loop^{n}(A, a))\).
From Section \ref{org6d38184},
we get a bijection \(\pi_{0}(A) \to \pi_{0}(\widetilde{A})\).
We can extend this result to all homotopy groups.
We show that \(i : A \to \widetilde{A}\) is \emph{\(\infty\)-connected}
in the sense that it induces a bijection between the \(n\)-th homotopy groups
for each \(n \ge 0\).

For a pointed types \((A, a) : \U_{\point}\),
\(\widetilde{A}\) has a base point
\(\widetilde{a} \equiv \lambda(X : \U, g : A \to X).ga\).
The maps \(i\) and \(j\) preserve base points,
and thus they induce maps
\(\Loop^{n}(i) : \Loop^{n}(A, a) \to \Loop^{n}(\widetilde{A}, \widetilde{a})\)
and \(\Loop^{n}(j) : \Loop^{n}(\widetilde{A}, \widetilde{a}) \to \Loop^{n}(A, a)\).
Identifying \(\Loop^{n}(\widetilde{A}, \widetilde{a})\) with
\(\prod_{X : \U}\prod_{g : A \to X}\Loop^{n}(X, ga)\)
by the functional extensionality,
we get:
\begin{align*}
\Loop^{n}(i) & : \Loop^{n}(A, a) \to \prod_{X : \U}\prod_{g : A \to X}\Loop^{n}(X, ga) &
\Loop^{n}(i) & =
\lambda p.\lambda(X, g).gp \\
\Loop^{n}(j) & : \left(\prod_{X : \U}\prod_{g : A \to X}\Loop^{n}(X, ga)\right) \to \Loop^{n}(A, a) &
\Loop^{n}(j) & =
\lambda\varphi.\varphi_{A}(\id_{A}).
\end{align*}
Then we have \(\Loop^{n}(j) \circ \Loop^{n}(i) = \id_{\Loop^{n}(A, a)}\).
For a closed term \(t : \Loop^{n}(\widetilde{A}, \widetilde{a})\),
we can construct a closed term of type
\[\prod_{X : \U}\prod_{g : A \to X}
   (\Loop^{n}(i)(\Loop^{n}(j)t))g = tg\]
in a similar way to Section \ref{org6d38184}.
Thus we conclude that the map \(i\) induces a bijection
\[\pi_{n}(A, a) \to \pi_{n}(\widetilde{A}, \widetilde{a})\]
for each \(n \ge 0\).

\subsection{Free Theorems for Open Terms}
\label{sec:orgb12e158}

In the reflexive graph model of Atkey et al. \cite{atkey2014relationally},
free theorems can be derived not only for closed terms
but also for open terms.
In our framework,
we cannot derive free theorems for open terms in general.
Indeed, the negation of the free theorem for some open term is provable.

Assuming the law of excluded middle for propositions in a universe \(\U\),
one can construct a function \(t : \prod_{X : \U}X \to X\)
such that \(t_{\Two}(0) = 1\) and \(t_{\Two}(1) = 0\)
where \(0 : \Two\) and \(1 : \Two\) are the constructors
of the two point type \(\Two : \U\) \cite[Exercise 6.9]{hottbook}.
Note that recently Booij et al. has pointed out that, conversely,
the existence of a non-trivial polymorphic endofunction
implies the law of excluded middle \cite{booij2017parametricity}.
Since the law of excluded middle for propositions in \(\U\)
can be expressed by some closed type \(\LEM_{\U}\),
\(t\) can be regarded as an open term
\[l : \LEM_{\U} \vdash t : \prod_{X : \U}X \to X.\]
For this open term the free theorem
\[\prod_{X, X' : \U}\prod_{f : X \to X'}\prod_{x : X}t(fx) = f(tx)\]
fails by taking \(f \equiv \lambda x.0 : \Two \to \Two\).
Since the law of excluded middle is consistent,
the free theorem for \(t\) is not provable.

\section{Church Encodings of Spaces}
\label{sec:org414f0ac}
\label{orgbf2ea29}
\label{org2e6c3bb}
In Section \ref{org8f38bec},
for each type \(A : \U\) we have an \(\infty\)-connected map
\(i : A \to \widetilde{A}\)
where \(\widetilde{A} \equiv \prod_{X : \U}(A \to X) \to X\).
For a concrete (higher) inductive type \(A\),
using the recursion principle of \(A\)
we have the \emph{Church encoding of \(A\)}
in Martin-L\"{o}f type theory without univalence or higher inductive types.
If \(A\) has a base point \(a_{0} : A\),
The Church encoding of \(A\) is of the form
\[\prod_{X : \U}\prod_{x : X}F_{A}(X, x) \to X\]
and its \(n\)-th loop space is
\[\prod_{X : \U}\prod_{x : X}F_{A}(X, x) \to \Loop^{n}(X, x),\]
where \(F_{A}(X, x)\) is a type defined from \(X\) and \(x\)
using only dependent products, dependent sums and path spaces.
The Church encoding of a type \(A\)
suggests that we can \emph{construct} generators of homotopy groups of \(A\)
without univalence or higher inductive types,
although we need univalence and higher inductive types
to \emph{prove} that they are actually generators of homotopy groups.

In this section we describe Church encodings
of some higher inductive types.
We also define the Hopf map
and give a generator of the third homotopy group of \(2\)-sphere
as a polymorphic function.

\subsection{The Circle}
\label{sec:org6da250e}
The circle \(\Sph^{1}\) is a higher inductive type
generated by a point constructor \(\base_{1} : \Sph^{1}\)
and a path constructor \(\Sphloop_{1} : \base_{1} = \base_{1}\).
It has a recursion principle
\[(\Sph^{1} \to X) \simeq \sum_{x : X}x = x.\]
Therefore
\[\widetilde{\Sph^{1}} \simeq
   \prod_{X : \U}\prod_{x : X}x = x \to X.\]
The constructors are defined as polymorphic functions
\begin{align*}
\widetilde{\base_{1}} &\equiv \lambda (X, x, p).x :
\prod_{X : \U}\prod_{x : X}x = x \to X \\
\widetilde{\Sphloop_{1}} &\equiv \lambda (X, x, p).p :
\prod_{X : \U}\prod_{x : X}x = x \to x = x
\end{align*}

\subsection{Spheres}
\label{sec:orgdc79e5a}
For a natural number \(n\),
the \emph{\(n\)-sphere \(\Sph^{n}\)}
is a higher inductive type generated by
a point constructor \(\base_{n} : \Sph^{n}\)
and a path constructor \(\Sphloop_{n} : \Loop^{n}(\Sph^{n}, \base_{n})\).
We have
\[\widetilde{\Sph^{n}} \simeq
   \prod_{X : \U}\prod_{x : X}\Loop^{n}(X, x) \to X.\]
The constructors are defined as polymorphic functions
\begin{align*}
\widetilde{\base_{n}} &\equiv \lambda (X, x, p).x :
\prod_{X : \U}\prod_{x : X}\Loop^{n}(X, x) \to X \\
\widetilde{\Sphloop_{n}} &\equiv \lambda (X, x, p).p :
\prod_{X : \U}\prod_{x : X}\Loop^{n}(X, x) \to \Loop^{n}(X, x)
\end{align*}
The \(k\)-th loop space of \(\widetilde{\Sph^{n}}\)
is \(\prod_{X : \U}\prod_{x : X}\Loop^{n}(X, x) \to \Loop^{k}(X, x)\)
studied in Section \ref{orgb86f605}.

\subsection{Suspensions}
\label{sec:orgeacf278}
For a type \(A\), the \emph{suspension \(\suspension A\) of \(A\)}
is a higher inductive type generated by
point constructors \(\suspN : \suspension A\) and \(\suspS : \suspension A\)
and a path constructor \(\glue : A \to \suspN = \suspS\).
We have
\[\widetilde{\suspension A} \simeq
   \prod_{X : \U}\prod_{x, y : X}(A \to x = y) \to X.\]

\subsection{Joins}
\label{sec:orgdabddb4}
For types \(A\) and \(B\),
the \emph{join \(A \join B\) of \(A\) and \(B\)}
is a higher inductive type generated by
point constructors \(\inl : A \to A \join B\) and \(\inr : B \to A \join B\)
and a path constructor \(\glue : \prod_{a : A}\prod_{b : B}\inl(a) = \inr(b)\).
We have
\[\widetilde{A \join B} \simeq
   \prod_{X : \U}\prod_{s : A \to X}\prod_{t : B \to X}
   (\prod_{a : A}\prod_{b : B}sa = tb) \to X.\]

\subsection{The Hopf Map}
\label{sec:org0fbff01}
The \emph{Hopf map} is a function \(\Sph^{3} \to \Sph^{2}\)
whose fiber at the base point is \(\Sph^{1}\).
Identifying \(\Sph^{3} \simeq \Sph^{1} \join \Sph^{1}\)
and \(\Sph^{2} \simeq \suspension \Sph^{1}\),
the Hopf map \(h : \Sph^{1} \join \Sph^{1} \to \suspension \Sph^{1}\)
is defined as
\(h(\inl(x)) \equiv \suspN\),
\(h(\inr(y)) \equiv \suspS\),
and \(h(\glue(x, y) = \glue(h_{1}(x, y))\),
where
\(h_{1} : \Sph^{1} \to \Sph^{1} \to \Sph^{1}\)
is a function defined as
\(h_{1}(\base_{1}, y) \equiv y\),
\(h_{1}(\Sphloop_{1}, \base_{1}) = \Sphloop_{1}^{-1}\),
and \(h_{1}(\Sphloop_{1}, \Sphloop_{1})\) is given by a proof of
\(\Sphloop_{1}^{-1} \concat \Sphloop_{1} = \refl_{\base_{1}} = \Sphloop_{1} \concat \Sphloop_{1}^{-1}\).

We define the Hopf map as a polymorphic function.
Observe that
\begin{align*}
& \Sph^{1} \join \Sph^{1} \to X 
\simeq \sum_{f, g : \Sph^{1} \to X}
\prod_{x, y : \Sph^{1}}fx = gy \\
\simeq & \sum_{x : X}\sum_{l : x = x}\sum_{y : X}\sum_{k : y = y}
\sum_{p : x = y}\sum_{\alpha : l \concat p = p}
\sum_{\beta : p \concat k = p}(\alpha \whisk{r} k) \concat \beta = (l \whisk{l} \beta) \concat \alpha
\end{align*}
and
\[\suspension \Sph^{1} \to X
   \simeq \sum_{x, y : X}\Sph^{1} \to x = y
   \simeq \sum_{x, y : X}\sum_{p : x = y}p = p.\]
Then we define
\begin{align*}
h : {} & (\prod_{X : \U}\prod_{x, y : X}\prod_{l : x = x}\prod_{k : y = y}
\prod_{p : x = y}\prod_{\alpha : l \concat p = p}\prod_{\beta : p \concat k = p}
(\alpha \whisk{r} k) \concat \beta = (l \whisk{l} \beta) \concat \alpha \to X) \\
& \to (\prod_{X : \U}\prod_{x, y : X}\prod_{p : x = y}p = p \to X) \\
h(f) \equiv {} & \lambda (X, x, y, p, \alpha).
f(X, x, y, \refl_{x}, \refl_{y}, p, \alpha^{-1}, \alpha, \check{\alpha})
\end{align*}
where \(\check{\alpha}\) is a proof of \(\alpha^{-1} \concat \alpha = \refl_{p} = \alpha \concat \alpha^{-1}\).

\subsection{A Generator of \(\pi_{3}(\Sph^{2})\)}
\label{sec:org04cab12}
The Hopf map is a generator of \(\pi_{3}(\Sph^{2})\).
We describe the generator as a polymorphic function.

First we define a \(3\)-loop of \(\Sph^{1} \join \Sph^{1}\)
as a polymorphic function.
We have to construct a function
\[l_{3} : {} \prod_{X : \U}\prod_{x, y :X}\prod_{l : x = x}\prod_{k : y = y}
   \prod_{p : x = y}\prod_{\alpha : l \concat p = p}\prod_{\beta : p \concat k = p}
   (\alpha \whisk{r} k) \concat \beta = (l \whisk{l} \beta) \concat \alpha \to \Loop^{3}(X, x).\]
By path induction on \(p\),
we can assume \(y \equiv x\) and \(p \equiv \refl_{x}\).
Then the goal becomes
\[l'_{3} : \prod_{X : \U}\prod_{x : X}\prod_{l, k : x = x}
   \prod_{\alpha : l = \refl_{x}}\prod_{\beta : k = \refl_{x}}
   (\alpha \whisk{r} k) \concat \beta = (l \whisk{l} \beta) \concat \alpha \to \Loop^{3}(X, x).\]
For \(\sigma : (\alpha \whisk{r} k) \concat \beta = (l \whisk{l} \beta) \concat \alpha\),
define \(l'_{3}(\sigma) : \refl^{2}_{x} = \refl^{2}_{x}\)
as the following concatenation:
\[\refl^{2}_{x}
   = (\beta^{-1} \concat (\alpha \whisk{r} k)^{-1})
   \concat ((\alpha \whisk{r} k) \concat \beta)
   \overset{E, \sigma}{=} (\alpha^{-1} \concat (l \whisk{l} \beta)^{-1})
   \concat ((l \whisk{l} \beta) \concat \alpha)
   = \refl^{2}_{x}\]
where \(E \equiv E(\alpha, \beta) : \beta^{-1} \concat (\alpha \whisk{r} k)^{-1}
   = \alpha^{-1} \concat (l \whisk{l} \beta)^{-1}\)
is the path described in Figure \ref{orgdc914d2},
also defined as \(E(\refl^{2}_{x}, \refl^{2}_{x}) \equiv \refl^{3}_{x}\)
by path induction on \(\alpha\) and \(\beta\).
\begin{figure}
\[\begin{tikzcd}
   x \arrow[-,r,"k"'] \arrow[-,rr,"\refl_{x}",bend left=100,""{name=r0,below}]
   \arrow[equal,from=r0,r,"\beta"'] &
   x \arrow[-,r,"l"',""{name=r2,above}] \arrow[-,r,"\refl_{x}",bend left=50,""{name=r1,below}]
   \arrow[equal,from=r1,to=r2,"\alpha"] & x
   \end{tikzcd} = \begin{tikzcd}
   x \arrow[-,r,"\refl_{x}",bend left=60,""{name=r0,below}] \arrow[-,r,"k"',""{name=r1,above}]
   \arrow[equal,from=r0,to=r1,"\beta"] &
   x \arrow[-,r,"\refl_{x}",bend left=60,""{name=s0,below}] \arrow[-,r,"l"',""{name=s1,above}]
   \arrow[equal,from=s0,to=s1,"\alpha"] & x
   \end{tikzcd} = \begin{tikzcd}
   x \arrow[-,rr,"\refl_{x}",bend left=100,""{name=r2,below}]
   \arrow[-,r,"\refl_{x}",bend left=50,""{name=r0,below}]
   \arrow[-,r,"k"',""{name=r1,above}] \arrow[equal,from=r0,to=r1,"\beta"] &
   x \arrow[-,r,"l"'] \arrow[equal,from=r2,"\alpha"] & x
   \end{tikzcd}\]
\caption{\label{orgdc914d2}
\(E : \beta^{-1} \concat (\alpha \whisk{r} k)^{-1} = \alpha^{-1} \concat (l \whisk{l} \beta)^{-1}\)}
\end{figure}

Now we can define a \(3\)-loop of \(\suspension \Sph^{1}\)
in a similar way to the Hopf map:
\begin{align*}
c & : \prod_{X : \U}\prod_{x, y : X}\prod_{p : x = y}p = p \to \Loop^{3}(X, x) \\
c & \equiv \lambda (X, x, y, p, \alpha).
l_{3}(X, x, y, \refl_{x}, \refl_{y}, p, \alpha^{-1}, \alpha, \check{\alpha}).
\end{align*}
We can also define it as an element of \(\Loop^{3}(\Sph^{2})\):
\begin{align*}
c & : \prod_{X : \U}\prod_{x : X}\Loop^{2}(X, x) \to \Loop^{3}(X, x) \\
c & \equiv \lambda (X, x, \alpha).
l_{3}(X, x, x, \refl_{x}, \refl_{x}, \refl_{x}, \alpha^{-1}, \alpha, \check{\alpha}).
\end{align*}
In fact \(c(\alpha)\) is the concatenation of paths
\[\refl^{2}_{x} = \alpha \concat \alpha^{-1}
   \overset{E}{=} \alpha^{-1} \concat \alpha = \refl^{2}_{x}\]
where \(E\) comes from the commutativity of concatenation of higher loops.

Here is a natural question.

\begin{question}
Is any generator of a homotopy group of a space
definable as a polymorphic function without univalence or higher inductive types?
\end{question}

This question is important because
it measures power of univalence and higher inductive types.
If the answer to the question is yes,
we can say, informally, that univalence and higher inductive types
give proofs that some elements are different
but do not generate new elements,
although there is a problem
which terms we should think of as proofs,
because in dependent type theory elements and proofs are not distinguished.

\section{Homotopy Type Theory}
\label{sec:org09dfe1d}
\label{org2b91024}
In the rest of this paper we prove the abstraction theorem.
We begin with a quick review of homotopy type theory \cite{hottbook}.

In this paper we consider the Martin-L\"{o}f's dependent type theory
\(\theory{T}\) with countably many univalent universes
\[\U_{0} : \U_{1} : \U_{2} : \dots,\]
an empty type \(\Zero : \U_{0}\),
a one point type \(\One : \U_{0}\),
a two point type \(\Two : \U_{0}\),
indexed \(\W\)-types \(\W[t, A, B]\)
and \(n\)-spheres \(\Sph^{n} : \U_{0}\).
The existence of ordinary \(\W\)-types is not enough to construct \(\W\)-types in the relational model,
and we require \emph{indexed \(\W\)-types}.
In extensional type theory
the existence of \(\W\)-types implies the existence of indexed \(\W\)-types
\cite{gambino2004wellfounded},
but in intensional type theory this does not hold
due to the lack of equalizers.
Also to construct general higher inductive types in the relational model
we need some class of indexed higher inductive types,
but we do not know such a class of higher inductive types.
Therefore we deal with only constant higher inductive types \(\Sph^{n}\).

For a type family \(i : I, x : A(i) \vdash B(x) \ \type\)
and a function \(i : I, x : A(i), y : B(x) \vdash t(y) : I\),
the \emph{\(\W\)-type \(\W[t, A, B]\) of \(B\) on \(A\) indexed over \(t\)}
is an inductive type family \(i : I \vdash \W[t, A, B](i) \ \type\)
with a single constructor
\[i : I, a : A(i), f : \prod_{y : B(a)}\W[t, A, B](ty)
  \vdash \Wsup_{[t, A, B]}(a, f) : \W[t, A, B](i).\]
We often omit the subscript \(_{[t, A, B]}\) of the constructor and
write it simply as \(\Wsup\).
The indexed \(\W\)-type has an induction principle:
given a type family \(i : I, w : \W[t, A, B](i) \vdash D(w) \ \type\)
and a term
\[i : I, a : A(i), f : \textstyle\prod_{y : B(a)}\W[t, A, B](ty),
  g : \textstyle\prod_{y : B(a)}D(fy) \vdash d(a, f, g) : D(\Wsup(a, f)),\]
we get a term
\[i : I, w : \W[t, A, B](i) \vdash \ind_{\W[t, A, B]}^{D}(d, w) : D(w)\]
together with a computational rule
\[\ind_{\W[t, A, B]}^{D}(d, \Wsup(a, f)) \equiv d(a, f, \lambda(y : B(a)).\ind_{\W[t, A, B]}^{D}(d, fy)).\]
There are projections
\begin{align*}
& i : I \vdash \proj_{1} : \W[t, A, B](i) \to A(i) \\
& i : I \vdash \proj_{1}(\Wsup(a, f)) \equiv a \\
& i : I \vdash \proj_{2} : \prod_{w : \W[t, A, B](i)}\prod_{y : B(\proj_{1}(w))}\W[t, A, B](ty) \\
& i : I \vdash \proj_{2}(\Wsup(a, f)) \equiv f.
\end{align*}

Some important types are definable from these types.
Ordinary \(\W\)-types \(\W_{x : A}B(x)\) are \(\W\)-types indexed over
the function \(B \to \One\).
The type \(\N\) of natural numbers is defined as
\[\N \equiv \W_{x : \Two}\rec_{\Two}(\Zero, \One, x),\]
where \(\rec_{\Two}(\Zero, \One) : \Two \to \U_{0}\) is a function defined by recursion
as \(\rec_{\Two}(\Zero, \One, 0_{\Two}) \equiv \Zero\) and
\(\rec_{\Two}(\Zero, \One, 1_{\Two}) \equiv \One\).
A coproduct \(A + B\) of two types \(A, B : \U\) is defined as
\[A + B \equiv \sum_{x : \Two}\rec_{\Two}(A, B, x).\]

For a function \(f : A \to B\), define
\[\isequiv(f) \equiv \left(\sum_{g : B \to A}\prod_{a : A}g(fa) = a\right)
  \times \left(\sum_{h : B \to A}\prod_{b : B}f(hb) = b\right)\]
and \((A \simeq B) \equiv \sum_{f : A \to B}\isequiv(f)\).
For types \(A, B : \U\), define a function \(\idtoequiv_{A, B} : A = B \to A \simeq B\) by path induction
as \(\idtoequiv(\refl_{A})\) is the identity function on \(A\).
The \emph{univalence axiom} is the axiom that \(\idtoequiv\) is an equivalence:
\[\ua_{\U} : \prod_{A, B : \U}\isequiv(\idtoequiv_{A, B}).\]

\section{Relational Model}
\label{sec:org3a2affd}
\label{org103b3d7}
The key to prove the abstraction theorem is the fact that
binary type families \(x : A, x' : A' \vdash \relation{A}(x, x') \ \type\)
form a model \(\Rel(\theory{T})\) of homotopy type theory
which we call the \emph{relational model}.

Families \(\related{x} : \relation{A} \vdash \relation{B}(\related{x}) \ \rel\)
of binary type families
are defined in Section \ref{org72cd84a}.
A \emph{term} of a family \(\related{x} : \relation{A} \vdash \relation{B}(\related{x}) \ \rel\)
is a triple of terms
\(x : A \vdash b(x) : B(x)\), \(x' : A' \vdash b'(x') : B'(x')\) and
\(x : A, x' : A', \related{x} : \relation{A}(x, x') \vdash
  \related{b}(\related{x}) : \relation{B}(\related{x}, b, b')\),
written \(\related{x} : \relation{A} \vdash \related{b}(\related{x}) : \relation{B}(\related{x})\)
in short.
In Section \ref{org72cd84a},
we defined dependent products, dependent sums,
path spaces and universes of binary type families.
It remains to construct other types and check the univalence axiom.

For a type constant \(C \equiv \Zero, \One, \Two, \Sph^{n}\),
the binary type family \(c : C, c' : C \vdash c = c' \ \type\)
has the same constructors and
satisfies the same induction principle as those of \(C\).
For example, \(c : \Two, c' : \Two \vdash c = c' \ \type\)
has two constructors \((0_{\Two}, 0_{\Two}, \refl_{0_{\Two}})\)
and \((1_{\Two}, 1_{\Two}, \refl_{1_{\Two}})\).
To see the induction principle of two point type,
let \(c : \Two, c' : \Two, \related{c} : c = c', x : A(c), x' : A'(c') \vdash
  \relation{A}(\related{c}, x, x') \ \type\)
be a family on \(c = c'\) and
\((a_{0} : A(0_{\Two}), a'_{0} : A'(0_{\Two}), \related{a}_{0} : \relation{A}(\refl_{0_{\Two}}, a_{0}, a'_{0}))\)
and \((a_{1} : A(1_{\Two}), a'_{1} : A'(1_{\Two}), \related{a}_{1} : \relation{A}(\refl_{1_{\Two}}, a_{1}, a'_{1}))\)
be elements of \(\relation{A}\).
We have to construct terms
\(c : \Two \vdash f(c) : A(c)\),
\(c' : \Two \vdash f'(c') : A'(c')\) and
\(c : \Two, c' : \Two, \related{c} : c = c' \vdash \related{f}(\related{c}) : \relation{A}(\related{c}, f(c), f'(c'))\)
such that \(f(0_{\Two}) \equiv a_{0}\), \(f(1_{\Two}) \equiv a_{1}\),
\(f'(0_{\Two}) \equiv a'_{0}\), \(f'(1_{\Two}) \equiv a'_{1}\),
\(\related{f}(\refl_{0_{\Two}}) \equiv \related{a}_{0}\) and
\(\related{f}(\refl_{1_{\Two}}) \equiv \related{a}_{1}\).
Define \(f\) and \(f'\) by \(\Two\)-induction.
By path induction,
to construct \(\related{f}\) it suffices to give a term
\(c : \Two \vdash \related{f}(\refl_{c}) : \relation{A}(\refl_{c}, f(c), f'(c))\),
which is given by \(\Two\)-induction.

To define indexed \(\W\)-types,
suppose that we get a family of binary type families
\(\related{i} : \relation{I}, \related{x} : \relation{A}(\related{i})
  \vdash \relation{B}(\related{x}) \ \rel\)
and a term
\(\related{i} : \relation{I}, \related{x} : \relation{A}(\related{i}),
  \related{y} : \relation{B}(\related{x}) \vdash \related{t}(\related{y}) : \relation{I}\).
First we have indexed \(\W\)-types
\(i : I \vdash \W[t, A, B](i) \ \type\)
and \(i' : I' \vdash \W[t', A', B'](i') \ \type\)
which we refer to as \(W(i)\) and \(W'(i')\) respectively.
We have to construct a type
\(i : I, i' : I', \related{i} : \relation{I}(i, i'),
  w : W(i), w' : W'(i') \vdash \relation{W}(\related{i}, w, w') \ \type\).
Let \(J \equiv \sum_{i : I}\sum_{i' : I'}\relation{I}(i, i') \times W(i) \times W'(i')\).
Define type families \(j : J \vdash \check{A}(j) \ \type\) and
\(j : J, x : \check{A}(j) \vdash \check{B}(x) \ \type\) as
\begin{align*}
\check{A}(\related{i}, w, w') & \equiv \relation{A}(\related{i}, \proj_{1}(w), \proj_{1}(w')) \\
\check{B}((\related{i}, w, w'), \related{a}) & \equiv \sum_{b : B(\proj_{1}(w))}\sum_{b' : B'(\proj_{1}(w'))}\relation{B}(\related{a}, b, b').
\end{align*}
Define a term \(j : J, x : \check{A}(j), y : \check{B}(x) \vdash \check{t}(y) : J\) as
\[\check{t}((\related{i}, w, w'), \related{x}, (b, b', \related{b})) \equiv
  (t(b), t'(b'), \related{t}(\related{b}), \proj_{2}(w)(b), \proj_{2}(w')(b')).\]
Then we set
\[\relation{W}(\related{i}, w, w') \equiv
  \W[\check{t}, \check{A}, \check{B}](\related{i}, w, w').\]
We have a constructor
\begin{align*}
& i : I, i' : I', \related{i} : \relation{I}(i, i'),
a : A(i), a' : A'(i'), \related{a} : \relation{A}(\related{i}, a, a'), \\
& f : \prod_{y : B(a)}W(t(y)),
f' : \prod_{y' : B'(a')}W'(t'(y')),
\related{f} : \prod_{y : B(a)}\prod_{y' : B'(a')}\prod_{\related{y} : \relation{B}(\related{a}, y, y')}
\relation{W}(\related{t}(\related{y}), f(y), f'(y')) \\
& \vdash \Wsup_{[\check{t}, \check{A}, \check{B}]}(\related{a}, \related{f}) :
\relation{W}(\related{i}, \Wsup_{[t, A, B]}(a, f), \Wsup_{[t', A', B']}(a', f')).
\end{align*}
One can check the induction principle of indexed \(\W\)-type.
Note that we have formalized, in Agda\footnote{\url{http://wiki.portal.chalmers.se/agda/}},
the construction of indexed \(\W\)-types
in the relational model\footnote{\url{https://gist.github.com/uemurax/040d22a4c037f5323ed26fbee6420544}}.

We give a sketch of a proof that
a universe \(X : \U, X' : \U \vdash X \to X' \to \U \ \type\)
of binary type families satisfies the univalence axiom.
Recall that \(\U\) satisfies the univalence axiom if and only if
the canonical function
\begin{align*}
e & : \U \to \sum_{X, X' : \U}X \simeq X' &
e(X) & \equiv (X, X, \id_{X})
\end{align*}
is an equivalence.
Observe that in \(\Rel(\theory{T})\)
a function \(\related{f} : \relation{A} \to \relation{B}\)
is an equivalence if and only if
\(f : A \to B\) and \(f' : A' \to B'\) are equivalences and
\(\related{f}(x, x') : \relation{A}(x, x') \to \relation{B}(fx, f'x')\)
is an equivalence for all \(x : A\) and \(x' : A'\).
Therefore, to show that \(X \to X' \to \U\) is univalent,
it suffices to see that
\[\related{e} : (X \to X' \to \U) \to \sum_{\relation{X}, \relation{Y} : X \to X' \to \U}
  \prod_{x : X}\prod_{x' : X'}\relation{X}(x, x') \simeq \relation{Y}(x, x')\]
is an equivalence for all \(X, X' : \U\).
There is an equivalence
\[\left(\sum_{\relation{X}, \relation{Y} : X \to X' \to \U}
  \prod_{x : X}\prod_{x' : X'}\relation{X}(x, x') \simeq \relation{Y}(x, x')\right)
  \simeq \left(X \to X' \to \sum_{\relation{X}, \relation{Y} : \U}\relation{X} \simeq \relation{Y}\right),\]
and \(\related{e}\) is homotopic to
\[(X \to X' \to e) : (X \to X' \to \U) \to
  (X \to X' \to \sum_{\relation{X}, \relation{Y} : \U}\relation{X} \simeq \relation{Y})\]
along this equivalence.
The function \((X \to X' \to e)\) is an equivalence
by univalency of \(\U\).

\section{The Abstraction Theorem}
\label{sec:orgc4d896a}
\label{org8cf691c}
In Section \ref{org103b3d7}
we see that the binary type families form a model \(\Rel(\theory{T})\)
of Martin-L\"{o}f's dependent type theory
with countable univalent universes, an empty type,
a one point type, a two point type, indexed \(\W\)-type and \(n\)-spheres.
Thus we have an interpretation \(\itpr{-} : \theory{T} \to \Rel(\theory{T})\).
\(\itpr{-}\) takes a type judgment \(x : X \vdash A(x) \ \type\)
to a type judgment
\[x : X, x' : X, \related{x} : \itpr{X},
  a : A(x), a' : A(x') \vdash \itpr{A}(\related{x}, a, a') \ \type\]
and a term judgment \(x : X \vdash t(x) : A(x)\)
to a term judgment
\[x : X, x' : X, \related{x} : \itpr{X}
  \vdash \itpr{t}(\related{x}) : \itpr{A}(\related{x}, t(x), t(x')).\]
Now the abstraction theorem is proved by taking \(\hat{t} \equiv \itpr{t}\).

\begin{theorem}[Abstraction Theorem]
For each term \(x : X \vdash t(x) : A(x)\),
there exists a term
\[x : X, x' : X', \related{x} : \itpr{X}(x, x') \vdash
  \hat{t}(\related{x}) : \itpr{A}(\related{x}, t(x), t(x')).\]
\end{theorem}

\label{sec-9}
\bibliography{my-references}
\end{document}